# STUDYING THERMODYNAMICS OF METASTABLE STATES
*UDC 536.7*

**Yuri Kornyushin**

Maître Jean Brunschvig Research Unit, Chalet Shalva, Randogne, CH-3975

**Abstract**. *Simple classical thermodynamic approach to the general description of metastable states is presented. It makes possible to calculate the explicit dependence of the Gibbs free energy on temperature, to calculate the heat capacity, the thermodynamic barrier, dividing metastable and more stable states, and the thermal expansion coefficient. Thermodynamic stability under mechanical loading is considered. The influence of the heating (cooling) rate on the measured dynamic heat capacity is investigated. A phase shift of the temperature oscillations of an ac heated sample is shown to be determined by the relaxation time of the relaxation of the metastable nonequilibrium state back to the metastable equilibrium one. This dependence allows one to calculate the relaxation time. A general description of the metastable phase equilibrium is proposed. Metastable states in $AB_3$ alloys are considered. Reasons for the change from the diffusional mechanism of the supercritical nucleus growth to the martensitic one as the heating rate increases are discussed. The Ostwald stage rule is derived.*

## 1. INTRODUCTION

Metastable states are common ones in physics, nanophysics, materials science, and chemistry. Many amorphous materials (including some amorphous metals and alloys), superheated and supercooled phases are examples of the metastable states.

The problem of metastable states is related to the problem of phase transformations. The common feature is a transition from one phase to the other one. But what is different here is reversibility. Phase transformations are reversible or almost reversible changes. The second-order transition is completely reversible. The first-order transition is often reversible in a sense that it is possible to transform two phases, one into another, and reversely by means of cycling temperature. Direct and reverse first-order phase transitions almost never occur at the same temperature. Usually a low-temperature phase requires overheating to be transformed into a high-temperature one, and vice versa, the high-temperature phase requires overcooling at reverse transformation. This phenomenon is called hysteresis, and apart from this the first-order phase transitions are usually reversible. On the contrary, the majority of transitions from metastable states to more stable ones are irreversible. For example, a crystallization of amorphous materials and transformation of diamond to graphite are completely irreversible. The reason for the irreversibility is that in these cases the temperature of the equilibrium of the two phases does not exist. The metastable phase just collapses into more stable one. The transition is one from the phase with the higher Gibbs free energy (GFE) to that with the lower one. The jump occurs not in the first derivatives of the GFE (enthalpy and volume) like in the case of the first-order phase transition, but in the GFE itself. So this type of transitions could be called zero-order transitions (in the spirit of the P. Ehrenfest classification).

The aim of this paper is to formulate a classical thermodynamic description to investigate the limits of a relative stability of metastable states and their thermodynamic properties. The GFE will be expanded in a series of deviations of entropy, but not in the vicinity of the phase equilibrium (which does not exist in a general case), but in the vicinity of some starting point.

The author had started the classical thermodynamic description of metastable states in 1985 [1]. Let us consider it in detail here.

The GFE is defined as a function of thermodynamic variables $T$ (temperature) and $P$ (pressure) as a rule [2].

Let us consider now the GFE per atom, divided by Boltzmann constant, $k$, as a function of entropy (per atom and divided by $k$) $s$, temperature $T$, and pressure $P$ (with thermodynamic variables $s$ and $P$):

$$\phi(s,P) = h(s,P) - Ts, \qquad (1)$$

where $h(s,P)$ is the enthalpy per atom and divided by $k$.

Enthalpy $h$ is originally defined as a function of two thermodynamic variables: entropy $s$, and pressure $P$ [2]. That is $h = h(s,P)$. In a state of equilibrium enthalpy depends on $T$ via $s(T,P)$ only. To prove this let us assume the contrary: $h = h(s(T,P),T,P)$. Then at constant pressure the heat capacity of a body (per atom and divided by $k$), $c_P = (dh/dT)_P = (\partial h/\partial s)_P (\partial s/\partial T)_P + (\partial h/\partial T)_P$. As in the state of equilibrium $(\partial h/\partial s)_P = T$ [2], and $(\partial s/\partial T)_P = c_P/T$ [2], we have: $c_P = c_P + (\partial h/\partial T)_P$. From this follows that $(\partial h/\partial T)_P = 0$, that is $h = h(S,P)$ and not $h = h(S,T,P)$.

The GFE as a function of thermodynamic variables $s$ and $P$ has a remarkable property: in a state of equilibrium it has a minimum on $s$. Really, at fixed $T$ and $P$ $(\partial \phi/\partial s)_P = (\partial h/\partial s)_P - T$. In the state of equilibrium the right-hand part of this equation is zero: $(\partial h/\partial s)_P - T = 0$ [2]. So the first condition of a minimum, $(\partial \phi/\partial s)_P = 0$, is fulfilled. Now let us take into account that at fixed $T$ and $P$ $(\partial^2 \phi/\partial s^2)_P \equiv (\partial^2 h/\partial s^2)_P$. According to [2] in a state of equilibrium $(\partial^2 h/\partial s^2)_P = (T/c_P) > 0$. So the second condition of a minimum, $(\partial^2 \phi/\partial s^2)_P > 0$ is fulfilled also.

Here like in usual classical thermodynamics we talk about differentiation of thermodynamic functions like GFE, enthalpy, etc. This means that it is assumed implicitly that these functions are defined well enough not only in the point of equilibrium, but in some vicinity of it also. The same was assumed in a "configurational" model [1,3].

When the GFE is regarded as a function of thermodynamic variables $s$ and $P$ with $T$ and $P$ being fixed, a minimum of the GFE on $s$ corresponds to a state of equilibrium, and an equation of equilibrium, $(\partial h/\partial s)_P = T$, determines explicit dependence of $s$ on $T$. This approach allows us to calculate explicit dependencies of the GFE, heat capacity and other thermodynamic quantities on $T$, and to formulate an approach for metastable states, which gives that thermodynamic barrier, dividing metastable and more stable states, is proportional to $(T_i - T)^{3/2}$ ($T_i$ is the temperature of the absolute instability). Thermodynamic instability under mechanical loading is considered also.

## 2. APPROXIMATION

Dependence of $h$ on $s$ is different for various systems. This problem is discussed in detail in [3]. For a metastable state, $h$ has a minimum at some value of $s$, and at some other value of $s$ it should have a maximum. The described barrier is a main feature of a metastable state. Here let us restrict ourselves with the case when $h = h(s,P)$ can be expanded in power series in a small parameter, $(s - s_0)$ ($s_0$ is the entropy of a system at $T = 0$; for stable systems $s_0 = 0$). A case of an expansion about a non-zero-point entropy will be discussed later, in Sections 10 - 13 [see, e.g., Eq. (36)].

As in this section $(\partial h/\partial s)_P = T = 0$ at $s = s_0$, the first three terms of the series are [1]

$$h(s,P) = h(s_0,P) + (T_0/2)(s - s_0)^2 - (T_0^2/12T_i)(s - s_0)^3, \qquad (2)$$

where $T_0$ and $T_i$ are some parameters, depending on $P$.

When $(s - s_0)$ is not small, Eq. (2) can be regarded as a model, describing metastable states also, because it describes a situation when $h(s,P)$ goes through a minimum, and then a maximum with the increase in $(s - s_0)$, as it should be for a metastable state.

Such an approach is applicable for metastable systems, including disordered (e.g., amorphous [1]) ones, and for systems which include electrons of conductivity or disordered subsystems, e.g., disordered grain boundaries in polycrystals, randomly distributed dislocations or quenched vacancies in a crystal, solid solutions, etc.

## 3. THERMODYNAMIC STABILITY

Using Eq. (2) one can see that $\phi$ as a function of $(s - s_0)$ has a minimum at

$$s_{min} = s_0 + (2T_i/T_0)\{1 - [1 - (T/T_i)]^{1/2}\}, \qquad (3)$$

and a maximum at

$$s_{max} = s_0 + (2T_i/T_0)\{1 + [1 - (T/T_i)]^{1/2}\}. \qquad (4)$$

It will be shown later that in the frames of expansion into series (or model), Eq. (2), the parameter $T_i$ has a meaning of a virtual critical point (the temperature of the absolute instability) of a metastable phase.

Thermodynamic barrier (per atom and divided by $k$), separating metastable and more stable states, is described as

$$\Delta\phi(T) = \phi(s_{max}) - \phi(s_{min}) = (8/3)(T_i^{1/2}/T_0)(T_i - T)^{3/2}, \qquad (5)$$

and $s_{max} - s_{min} = (4T_i/T_0)[1 - (T/T_i)]^{1/2}$.

From Eq. (5) one can see that the thermodynamic barrier vanishes at $T \to T_i$. So $T_i$ has a meaning of the temperature of absolute instability. But metastable system cannot exist up to $T = T_i$ as the thermodynamic barrier becomes too small when $T$ approaches $T_i$. The system leaves metastable state when the temperature reaches some characteristic temperature, $T_c$, which has a meaning of a real critical temperature. It is often essentially smaller than $T_i$.

In this case real critical temperature could be approximated as [1]

$$T_c \approx (8/3)T_i^2/T_0. \tag{6}$$

System leaves metastable state for more stable one when $T$ is of the order of magnitude or larger than $T_c$, and accordingly to the kinetics [4] the higher the heating rate the higher the temperature at which this process has a considerable rate.

## 4. THERMODYNAMIC STABILITY UNDER CONDITIONS OF MECHANICAL LOADING

Thermodynamic stability under conditions of mechanical loading was first discussed in [5]. Proposed formalism allows us to obtain an expression for thermodynamic barrier [Eq. (5)]. When $T$ is essentially smaller than $T_i$ the thermodynamic barrier does not depend on $T$, but it depends on mechanical stresses. In general case both parameters in Eq. (5), $T_i$ and $T_0$, depend on stresses. As a result of this, the barrier itself depends on stresses also. Expansion of the barrier in power series in the tensor of stresses, $\sigma_{ik}$, for isotropic medium, yields

$$k\Delta\phi = k\Delta\phi(0) - (v/3)\sigma_{ii} - u_{el}, \tag{7}$$

where $\Delta\phi(0) = gT_c$ is $\Delta\phi$ at $\sigma_{ik} = 0$, $g$ is some coefficient of the order of unity, exact value of which depends on the rate of the change in thermodynamic conditions, such as $T_t/T$ ($T_t$ is the rate of the change of temperature), and on the temperature at which the instability occurs. Parameter $v$ represents the activation volume, $\sigma_{ii}$ is a spur of $\sigma_{ik}$ and $u_{el}$ is the elastic energy per one atom.

When $\Delta\phi \approx gT$, system leaves metastable state for more stable one. According to the given considerations and Eq. (5), the critical shear stress may be estimated as

$$P_c \approx [2kgG(T_c - T)/v_a]^{1/2}, \tag{8}$$

where $G$ is the shear modulus and $v_a$ is the volume per atom.

To estimate $P_c$ using Eq. (8) let us take $g = 1$, $v_a = 2\times10^{-23}$ cm$^3$, $G = 12\times10^9$ Pa, and $T_c - T = 370$ K. At such values of the parameters Eq. (8) yields $P_c = 2.48\times10^9$ Pa.

For the compression (elongation) we have

$$P_c \approx (|v|E/3v_a)[1 + 18kgv_a(T_c - T)/Ev^2]^{1/2} \pm Ev/3v_a, \tag{9}$$

where $E$ is the Young modulus, plus refers to compression and minus refers to elongation.

At $v \to 0$ Eq. (9) yields Eq. (8) where $G$ is replaced by $E$. At $T \to T_c$ Eq. (9) yields

$$P_c = \pm 3kg(T_c - T)/v, \tag{10}$$

where plus refers to elongation at $v > 0$ and minus refers to compression at $v < 0$.

In the case of compression and $v > 0$ as well as in the case of elongation and $v < 0$ we have respectively

$$P_c = \pm 2vE/3v_a. \tag{11}$$

Thermodynamic instability under loading may often result in a fracture because of the extreme brittleness or low strength of a more stable phase. This refers to the majority of the amorphous metallic alloys and other metastable phases, e.g., diamond (the stable phase, graphite, is of a very low strength).

## 5. THERMAL PROPERTIES

Heat capacity at constant pressure (per atom and divided by $k$) is given by [1]

$$c_P = T(\partial s/\partial T)_P = (T/T_0)[1 - (T/T_i)]^{-1/2}. \tag{12}$$

This equation is derived for the metastable equilibrium phase, so it is applicable when the phase is still stable, i.e. for $T < T_c$ only.

At $T \ll T_i$ $c_P = T/T_0$ which corresponds to a well-known result for amorphous materials [6].

At $T = T_i$ the isobaric heat capacity diverges.

Thermal expansion coefficient, $\alpha(T) = (\partial v_a/\partial T)_P/v_{a0} = -k(\partial s/\partial P)_T/v_{a0}$ ($v_a$ and $v_{a0}$ are the volume of the sample per atom at $T > 0$ and $T = 0$ respectively) is given by the formula [5]:

$$\alpha(T) = \alpha_0 + C(T/T_i)[1 - (T/T_i)]^{-1/2} + 2(D - C)\{1 - [1 - (T/T_i)]^{1/2}\}, \tag{13}$$

where

$$\alpha_0 = -k(\partial s_0/\partial P)/v_{a0}, \quad C = (k/v_{a0})(T_i/T_0)(\partial \ln T_i/\partial P) \text{ and } D = (k/v_{a0})(T_i/T_0)(\partial \ln T_0/\partial P). \tag{14}$$

At low temperatures, when $T$ is essentially smaller than $T_i$, we have:

$$\alpha(T) = \alpha_0 + D(T/T_i) + (C + D)(T/2T_i)^2 + (2C + D)(T/2T_i)^3 + \ldots . \tag{15}$$

Eq. (15) determines thermal expansion at low temperatures. Depending on values and signs of $C$ and $D$ there are three possible types of curves for $\alpha(T)$ at low $T$: monotonically increasing, having a maximum and then a minimum, and having only a minimum. At high $T$ the contribution of metastability may lead to increase in $\alpha(T)$ if $C < 0$. Such non-monotonic dependencies of $\alpha(T)$ are well known in amorphous solids [6], but the best known example of a system with a non-monotonic thermal expansion coefficient is water. It should be noted that in this case, thermal expansion anomalies occur already in a stable region (the results of this section are well applicable for a stable region also).

Thermodynamic properties, in particular, compressibility of supercooled water, heavy water and their mixtures were studied recently in [7,8].

The results on the thermal expansion coefficient are applicable for $T < T_c$ only, like in the case of the heat capacity.

Isothermal compressibility is known to be proportional to isobaric thermal expansion [2]. As the isobaric thermal expansion diverges at $T = T_i$, so the isothermal compressibility also diverges at $T = T_i$.

## 6. GENERAL CASE

Dependence of $h$ on $s$ is obviously different for various systems [5]. At low $T$ in a general case it may be approximated by the first term of expansion: $h(s,P) = A(P)(s - s_0)^{a(P)}$ ($A$ and $a$ are some constants, depending on pressure). Taking into account that $(\partial h/\partial s)_P = T$, we get that at $s = s_0$ $(\partial h/\partial s) = 0$. The only way to satisfy this condition is to restrict the range of the possible values of $a(P)$ : $a(P) > 1$. As we shall see later, the value of $a(P)$ determines the low-temperature heat capacity, which is linear in $T$ for conduction electrons [2], as well as for amorphous solids [6]. For such cases we have to adopt that the first term of expansion of $h$ is proportional to $(s - s_0)^2$. This case was considered above. Now let us consider a more general case, when the first two terms of expansion of the GFE on $(s - s_0)$ are arbitrary powers ($a$ and $b$, $1 < a < b$) of $(s - s_0)$. Such two terms give us the same type of dependence of $\phi(s,P)$ on $(s - s_0)$ as regarded in Section 2. That is, the dependence of $\phi$ as a function of $(s - s_0)$, having a minimum at some $s = s_{min}$ and then having a maximum at some $s = s_{max}$. From such type of dependence, as it will be shown later, it follows that there exists a virtual critical point (the temperature of the absolute instability) of a metastable phase. So the existence of the virtual critical point is not an assumption of the theory discussed here, but it is a consequence of the chosen type of dependence of $\phi$ on $(s - s_0)$. Only such dependence corresponds to a metastable state, which loses stability at higher temperatures.

In general case

$$\phi(s,P) = \phi(s_0,P) + [T_0/a(a-1)^{a-1}](s-s_0)^a - [T_0^{(b-1)/(a-1)}/T_i^{(b-a)/(a-1)}] \times$$
$$[(b-a)^{(b-a)/(a-1)}/b(a-1)^{b-2}(b-1)^{(b-1)/(a-1)}](s-s_0)^b - T(s-s_0), \quad (16)$$

where $b = b(P)$ is some constant, depending on pressure.

Condition of equilibrium, $(\partial \phi/\partial s)_P = 0$, yields equations for $s_{min}(T)$ and $s_{max}(T)$:

$$[T_0^{(b-1)/(a-1)}/T_i^{(b-a)/(a-1)}][(b-a)^{(b-a)/(a-1)}/(a-1)^{b-2}(b-1)^{(b-1)/(a-1)}](s-s_0)^{b-1} -$$
$$[T_0/(a-1)^{a-1}](s-s_0)^{a-1} + T = 0. \quad (17)$$

At $T \to 0$ $s_{min} \to s_0$, and Eq. (17) yields

$$s_{min} = s_0 + (a-1)(T/T_0)^{1/(a-1)}, \quad (18)$$

$$c_P = (T/T_0)^{1/(a-1)}. \quad (19)$$

At $T = 0$ $s_{min} = s_0$ and

$$s_{max} = s_0 + [(a-1)^{(b-a-1)/(b-a)}(b-1)^{(b-1)/(a-1)(b-a)}/(b-a)^{1/(a-1)}](T_i/T_0)^{1/(a-1)}. \quad (20)$$

Eqs. (16) and (20) allow to calculate the height of the thermodynamic barrier at $T = 0$:

$$(\Delta\phi)_{T=0} = [(b-1)^{a(b-1)/(a-1)(b-a)}/ab(a-1)^{(2a-b)(b-a)}(b-a)^{1/(a-1)}] \times$$
$$[T_i^{a/(a-1)}/T_0^{1/(a-1)}]. \quad (21)$$

At $a = 2$ and $b = 3$, $(\Delta\phi)_{T=0} = (8/3)(T_i^2/T_0)$ as expected.

At $T = T_i$ the thermodynamic barrier vanishes, $s_{min}(T_i) = s_{max}(T_i) = s_i$ and Eq. (17) yields

$$s_i = s_0 + (a-1)[(b-1)T_i/(b-a)T_0]^{1/(a-1)}. \qquad (22)$$

Now let us investigate the temperature dependence of the thermodynamic barrier near $T_c$ point. For this purpose let us expand $(\partial\phi/\partial s)_P$ in power series in $(s_i - s)$. The first two terms are

$$(\partial\phi/\partial s)_P = (T_i - T) - \{[(b-a)T_0]^{2/(a-1)}/2(a-1)[(b-1)T_i]^{(3-a)/(a-1)}\}(s_i - s)^2. \qquad (23)$$

Condition $(\partial\phi/\partial s)_P = 0$ and Eq. (23) yield

$$s_m = s_0 + (a-1)[(b-1)T_i/(b-a)T_0]^{1/(a-1)}\{1 \pm [2/(a-1)(b-1)]^{1/2}[1-(T/T_i)]^{1/2}\}, \qquad (24)$$

where $m$ is maximum or minimum and $+$ refers to the maximum and $-$ refers to the minimum.
   Eqs. (16) and (24) yield that in general case in the vicinity of $T_i$ the barrier,

$$\Delta\phi = (2^{5/2}/3)[T_i^{a/(a-1)}/T_0^{1/(a-1)}][(a-1)^{1/2}(b-1)^{(3-a)/2(a-1)}/(b-a)^{1/(a-1)}] \times$$
$$[1 - (T/T_i)]^{3/2}. \qquad (25)$$

Eq. (25) shows that in the vicinity of $T_i$ the exponent of the temperature dependence of the barrier in general case is 3/2, like in the simple case regarded above (the terms, proportional to $[1 - (T/T_i)]^{1/2}$ and to $[1 - (T/T_i)]$ annihilate). So the 3/2 rule seems to be the universal one.
   At $a = 2$ and $b = 3$ $\Delta\phi = (8/3)(T_i^2/T_0)[1 - (T/T_i)]^{3/2}$ as expected.

## 7. LIMITS OF THE VALIDITY OF THE THEORY

As follows from Eqs (3) and (4) the expansion performed in Eq. (2) is valid when $4T_i$ is essentially smaller than $T_0$ [5]. Both parameters, $T_i$ and $T_0$, ought to be taken from the comparison with the experimental results. To evaluate them one may use low $T$ heat capacity data. As an appropriate example one may take $T_0 = 50000$ K and $T_i = 3000$ K [5]. Then we have $T_c$ about 480 K, $s_{min} - s_0 \leq 0.12$ and $s_{max} - s_0 \leq 0.24$. As follows from Eq. (3), in general case the presented theory is valid when $T$ is essentially smaller than $T_i$ at any rate.
   As was mentioned above, Eq. (2) can be regarded as a model, describing metastable states also, because it describes a situation, when $h(s,P)$ goes through a minimum and then a maximum with the increase in $(s - s_0)$, as it should be for a metastable state.

## 8. HEAT CAPACITY AT A FINITE HEATING RATE

Heat capacity, $c$, is usually measured in finite heating (cooling) rate experiments. For the sake of simplicity let us consider a model of one relaxation time [5]. The relaxation time, $\tau$, may be considerable at low $T$ due to barriers, dividing different states of the system. These barriers may be penetrated by the way of tunneling. The rate of the relaxation of the enthalpy $h$ depends on the deviation of $h$ from its equilibrium value at a given $T$, $h_e(T)$. Let us consider small deviations, when it is possible to use a linear approximation [5]. In this case the kinetic equation is [5]

$$(dh/dt) + (h - h_e)/\tau = 0. \tag{26}$$

At low $T$ the equilibrium value of the heat capacity, $c_e = T/T_0$, $h_e = T^2/2T_0$. Eq. (26) yields

$$(dh/dT) + h/T_t\tau = T^2/2T_0T_t\tau. \tag{27}$$

Eq. (27) describes the evolution of $h = h(T)$ from the initial value, $h_0 = h(T_{in})$, at the beginning of the experiment. This equation is applicable only for low $T$. In this case $\tau$ usually does not depend on $T$.

At low heating rates, when $T_t\tau$ is considerably smaller than $T$, $h = h_e$ and $c = c_e$. At high heating rates, when $T_t\tau$ is considerably larger than $T$, relaxation is slow, $h$ is considerably smaller than $h_e$ and as follows from Eq. (27), $(dh/dT) = c = h_e/T_t\tau$. As an example let us consider a case when at $t = 0$ $T = 0$. In this case we have

$$h = (T^2/2T_0) - (T_t\tau/T_0)T + (T_t\tau)^2[1 - \exp(-T/T_t\tau)]/T_0. \tag{28}$$

Measured dynamic heat capacity is given by the following expression

$$c = (T/T_0) - (T_t\tau/T_0)[1 - \exp(-T/T_t\tau)]. \tag{29}$$

Eq. (29) shows that the dynamic heat capacity, $c < c_e$. At low heating rates, when $T_t\tau$ is considerably smaller than $T$, $c = T/T_0$. At high heating rates, when $T_t\tau$ is considerably larger than $T$, $c = T^2/2T_0T_t\tau$. Using this expression for more general case it is possible to estimate the relaxation time, $\tau$, and to use it to calculate the static heat capacity, $T/T_0$, with the aid of the measured heat capacity, $c$, and Eq. (29).

It is worthwhile to note that it was implicitly assumed in this section that the temperature $T$ remains homogeneous throughout the sample during heating with finite heating rate. This is possible only when the temperature equilibrates throughout the sample fast enough that is when the sample is thin enough [9]. Otherwise the characteristic time of the temperature equilibration appears in Eqs (26) - (29), instead of the relaxation time of the relaxation of the metastable nonequilibrium state back to the metastable equilibrium one.

The condition of the fast enough equilibration of the temperature may be difficult to fulfil. In this case another method to measure the relaxation time of the relaxation of the metastable nonequilibrium state back to the metastable equilibrium one can be applied. Let us consider this method in the following section [9].

## 9. PHASE SHIFT CAUSED BY THE RELAXATION

Let us consider a bulk sample with homogeneously distributed source of heat in the bulk of a sample [9]. The source can be the Joule heat. Let the sample be heated by ac, $I = I_0\sin\omega t$ ($I_0$ is the amplitude, and $\omega$ is the angular frequency). Then the Joule heat per one atom is

$$q(t) = q_e - q_e\cos 2\omega t, \tag{30}$$

where $q_e$ is the average Joule heat per atom divided by $k$.

Let us consider a sufficiently thick sample at a sort of isothermal regime. The heat power per atom and divided by $k$, being extracted out of the sample is $q_e$. Averaged over the volume of a sample and over the period of the oscillations, temperature, $\langle T \rangle$, does not depend on time. The temperature does not practically depend on coordinates throughout the bulk of a sample, and grad$T$ is essential in a thin surface layer only. When frequency is high enough, the heat exchange caused by the temperature oscillations inside the bulk of a sample is negligible and the energy conservation law in the bulk of a sample can be written as follows:

$$[dh(t)/dt] + q_e \cos 2\omega t = 0. \tag{31}$$

Eq. (26) and (31) yield

$$h(t) - h_e(T(t)) = q_e \tau \sin 2\omega t. \tag{32}$$

Let us regard a case when the deviation of the temperature from its averaged value $\langle T \rangle$ is small. Then in the linear approximation on this deviation

$$h_e(T(t)) = h_e(\langle T \rangle) + c_e(\langle T \rangle)[T(t) - \langle T \rangle]. \tag{33}$$

In this case Eqs (31) - (33) yield

$$T(t) = \langle T \rangle - (q_e \tau / c_e)\{[1 + (2\omega\tau)^2]^{1/2}/2\omega\tau\}\cos(2\omega t - \varphi), \tag{34}$$

$$\varphi = \arcsin[1 + (2\omega\tau)^2]^{-1/2}. \tag{35}$$

The phase shift, $\varphi$, decreases from $\pi/2$ to 0 when $\omega$ increases from zero to the range $2\omega\tau \gg 1$ concomitantly.

The best way to measure the phase shift is to take two samples of wire in series, one of them of a stable phase with zero or very small relaxation time (presumably a copper sample) and another one of a metastable state (e.g., some metallic glass sample) and to measure by some method the temperature oscillations in both samples. The difference in the phases of the oscillations in two samples allows one to calculate $\tau$, using Eq. (35).

## 10. METASTABLE HOMOGENEOUS-PHASE EQUILIBRIUM

Metastable states, arising during first-order phase transformations, is a thoroughly investigated branch [10]. Here let us see what could be gained using presented approach.

First-order phase transformations occur at temperatures different from the equilibrium of the phases temperature, $T_e$, for the phases with fixed compositions. The transition to higher-temperature phase requires overheating and that to the lower-temperature one, overcooling. The transition may involve various mechanisms dependent on the temperature variation rate. At low heating (cooling) rates the diffusional mechanism is a regular one for many of the first-order transformations. While at high rates the diffusional mechanism often changes to the martensitic one. The scope for these mechanisms is governed by the thermodynamic conditions in the superheated (supercooled) phase. It is governed by the height of the barrier, separating the state of an atom in equilibrium with the metastable phase from more stable state, as well as by the kinetic conditions: temperature variation rate, diffusion coefficients, etc. At higher rates of the temperature variation the overheating increases. This leads to the smaller barrier between the states of a phase, facilitating martensitic mechanism of transformation.

Let us start with a superheated metastable phase [11]. Let us expand the enthalpy in a power series in the deviation of $s$ from its equilibrium value, $s_e$, at $T = T_e$. The first four terms of the series are

$$h(s,P) = h(s_e,P) + T_e(s - s_e) + (T_0/2)(s - s_e)^2 - [T_0^2/12(T_i - T_e)](s - s_e)^3. \tag{36}$$

This expansion is valid for $(s - s_e) \ll 1$; in other cases it can be considered as a model representation for $h$ describing a system in a metastable state for

$$s = s_{min} = s_e + [2(T_i - T_e)/T_0]\{1 - [(T_i - T)/(T_i - T_e)]^{1/2}\}, \tag{37}$$

and in unstable equilibrium for

$$s = s_{max} = s_e + [2(T_i - T_e)/T_0]\{1 + [(T_i - T)/(T_i - T_e)]^{1/2}\}. \tag{38}$$

It is worthwhile to note that $s_{min} - s_e \leq 2(T_i - T_e)/T_0$, and $s_{max} - s_e \leq 4(T_i - T_e)/T_0$. So Eq. (36) is valid for $4(T_i - T_e) \ll T_0$.

The first-order transition of a diffusional type occurs by the more stable phase nucleating and the nuclei larger than the critical ones growing by atoms in equilibrium with the metastable phase passing to a state of the equilibrium with the atoms in the nuclei. An atom in equilibrium with the metastable phase is in a potential well, bottom of which is determined by $\phi(s_{min})$ and the crest by $\phi(s_{max})$, so the barrier height is described by

$$\Delta\phi(T) = \phi(s_{max}) - \phi(s_{min}) = (8/3)(T_i - T_e)^{1/2}(T_i^{3/2}/T_0)[1 - (T/T_i)]^{3/2}. \tag{39}$$

Eq. (39) takes into account only the barrier, separating the homogeneous metastable phase from more stable one. The contribution of the surface energy is not taken into account.

Eq. (39) shows that the height of the barrier tends to zero as $T$ approaches $T_i$ in accordance with the 3/2 law. The contribution of the surface energy is not important at that because the metastable phase itself becomes unstable.

Under the isothermal conditions the transformation can occur when $T$ is of the order of magnitude or larger than $\Delta\phi(T)$ because the individual atoms overcome the barrier and attach to the nuclei by diffusion if the kinetic conditions allow it. At high heating (cooling) rates there may be a substantial superheating (supercooling). As the rate increases, $T \to T_i$, and the barrier height decreases. At sufficiently high rates, the barrier height is reduced so much as to become unimportant and the martensitic growth replaces the diffusion-limited one.

Eqs (36) - (39) are valid for the supercooled phase also [9]. Only one should take into account that for the superheated phase $s_e \leq s$, $T_e \leq T \leq T_i$ and $s_{min} \leq s_{max}$, while for the supercooled phase $s \leq s_e$, $T_i \leq T \leq T_e$ and $s_{max} \leq s_{min}$.

## 11. METASTABLE-PHASE THERMAL PARAMETERS

The specific heat at the constant pressure [11],

$$c_P = T(\partial s/\partial T)_P = (T/T_0)[(T_i - T_e)/(T_i - T)]^{1/2}. \tag{40}$$

It has a square root singularity at $T_i$, which becomes more prominent as the rate increases and thus the actual transition temperature approaches $T_i$.

Let us consider now the thermal expansion coefficient [11]. The thermal expansion coefficient, $\alpha(T) = (\partial v_a/\partial T)_P/v_e = -k(\partial s/\partial P)_T/v_e$ ($v_a$ and $v_e$ are the volume of the sample per atom at $T > T_e$ and $T = T_e$ respectively) is given by the following equation:

$$\alpha(T) = \alpha_e + K[1 - (T/T_i)]^{1/2} + B[1 - (T/T_i)]^{-1/2}, \tag{41}$$

where

$$\alpha_e = -(k/v_e)\{(\partial s_e/\partial P) + 2[\partial((T_i - T_e)/T_0)/\partial P]\},$$
$$K = (k/v_e)\{2[T_i/(T_i - T_e)]^{1/2}[\partial((T_i - T_e)/T_0)/\partial P] - (1/T_0)[T_i/(T_i - T_e)]^{1/2}[\partial(T_i - T_e)/\partial P]\},$$
$$B = (k/v_e T_0)[(T_i - T_e)/T_i]^{1/2}(\partial T_i/\partial P). \tag{42}$$

At $T = T_e$

$$\alpha(T_e) = -(k/v_e)(\partial s_e/\partial P) + (k/v_e T_0)(\partial T_e/\partial P). \tag{43}$$

It is worthwhile to note that Eq. (40) and (41) yield well-known results. In [12] these results were obtained for a ferromagnet in the mean-field approximation. In [13] more general expressions that give the exponents characterizing the divergencies along several thermodynamic paths (isothermal, isobaric) were derived. These results are summarized in Chapter 2 of [14]. In this particular case nothing is gained by the author's alternative approach. It just shows that the approach discussed here yields correct results in this case also. But it should be mentioned that the accepted framework allows one to consider a series of other problems.

Eq. (41) shows that $\alpha(T)$ has a square-root singularity (like $c_P$) as $T$ approaches $T_i$. But the results, obtained in this section are derived for the metastable equilibrium phase, so they are applicable when the phase is still stable, i.e. for $T < T_c$ only.

## 12. METASTABLE STATES IN $AB_3$ ALLOYS

Let us consider a first-order phase transition in an ordering alloy having $AB_3$ stoichiometry [15,11]. For $T_e < T < T_i$, a superheated ordered phase exists in a metastable state, as is evident from the $AB_3$ phase diagram, which has been examined by numerical methods in the Gorsky-Bragg-Williams approximation [15]. The part of the free energy, $F$, dependent on the order parameter, $\eta$, is [15]

$$F = -0.457 NkT_e \eta^2 + 0.0625 NkTf(\eta), \tag{44}$$

where $N$ is the total number of atoms in the alloy and

$$f(\eta) = 3(1 - \eta)\ln[3(1 - \eta)^2/16] + (1 + 3\eta)\ln[(1 + 3\eta)/4] + 3(3 + \eta)\ln[(3 + \eta)/4]. \tag{45}$$

It is possible to introduce a free parameter (another order parameter), $x$, as follows:

$$x = [f(\eta_e) - f(\eta)]/16, \qquad (46)$$

where $\eta_e = 0.463$ is the equilibrium value of $\eta$ in the ordered phase at $T = T_e$.

For small $x$ the last three equations yield

$$F = Nk[T_e(x + 0.205x^2 - 1.04x^3) - Tx]. \qquad (47)$$

Eq. (47) corresponds exactly to the initial equations in the general theory, Eqs (1) and (2), with $T_0 = 0.410T_e$ and $T_i = 1.0134T_e$, so the expression for the free energy of $AB_3$ alloy in the discussed model is reduced to the initial expression in the general theory by changing variables. The values of $x$ corresponding to the minimum do not exceed 0.0656 throughout the metastable-state range, which justifies the $x$ expansion used. The values of $x$ corresponding to the maximum do not exceed 0.131.

The usual value of $T_e$ for $AB_3$ alloys is hundreds of degrees of Kelvin, so the metastable-range temperature width is up to 10 degrees.

In the $AB_3$ systems $\Delta\phi$ is very small. The maximum value of it at $T = T_e$ is $0.00118T_e$, as Eq. (39) shows, so there is virtually no barrier and a critical nucleus grows very rapidly, which agrees with experiment [15].

## 13. SOME REMARKS ON SECOND-ORDER PHASE TRANSITIONS

The enthalpy for the second-order phase transition should be written as follows [9]:

$$h(s,P) = h(s_e,P) + T_e(s - s_e) + [(1 - \beta)^{(2-\beta)/(1-\beta)}/(2 - \beta)]T_0(s_e - s)^{(2-\beta)/(1-\beta)} + \dots . \qquad (48)$$

This equation has been written for $T \leq T_e$, that is for $s \leq s_e$. Eq. (48) and $(\partial\phi/\partial s) = 0$ yield

$$s = s_e - [1/(1 - \beta)][(T_e - T)/T_0]^{1-\beta}. \qquad (49)$$

As $c_P = T(\partial s/\partial T)_P$, Eq. (49) yields for the heat capacity and the thermal expansion coefficient the following results [9]

$$c_P = (T/T_0)[(T_e - T)/T_0]^{-\beta},$$
$$\alpha(T) = -(k/v_e)\{(\partial s_e/\partial P) + (\partial \ln T_0/\partial P)[(T_e - T)/T_0]^{1-\beta} - (1/T_0)(\partial T_e/\partial P)[(T_e - T)/T_0]^{-\beta}\}. \qquad (50)$$

These equations show that $\beta$ has the meaning of a critical index. The equation for $\alpha$ has been derived on the assumption that $\beta$ does not depend on pressure.

## 14. DISCUSSION

Classical thermodynamic method, described above, gives explicit $T$ relationships for the barrier separating the states, the specific heat and the thermal expansion coefficient. This approach may be useful in other cases. For example if there is a set of the intermediate metastable phases between the two major ones, which correspond to minima in $\phi$ with respect to $s$, in which case slow uniform heating with restricted entropy increase causes the sample to enter the corresponding minima in order of increasing entropy (and in the reverse order during cooling), i.e. in a sequence, corresponding to the Ostwald stage rule [11].

The approach may be useful in other cases also. The metastable states in $AB_3$ alloys are considered and the same approach is applied to the second-order transformations.

For the dielectric materials low temperature heat capacity is determined by phonons and it is proportional to $T^3$ (the Debye $T^3$ law) [16]. This corresponds to $a = 4/3$ in Section 6. For $b = 5/3$ the problem can be solved analytically [9]. In this case [9]

$$s_m = s_0 + (8/3)(T_i/T_0)^3 \{1 \pm [1 - (T/T_i)]^{1/2}\}^3, \tag{51}$$

where $m$ is maximum or minimum and $+$ refers to the maximum and $-$ refers to the minimum.

The heat capacity in this case [9],

$$c_P = (4T_i^2/T_0^3)\{1 - [1 - (T/T_i)]^{1/2}\}^2 T/[1 - (T/T_i)]^{1/2}, \tag{52}$$

and the thermodynamic barrier is as follows [9]:

$$\Delta\phi(T) = (32/3)(T_i^4/T_0^3)\{[1 - (T/T_i)]^{3/2} + 0.2[1 - (T/T_i)]^{5/2}\}. \tag{53}$$

Eq. (53) shows that in the vicinity of $T_i$ the barrier is proportional to $[1 - (T/T_i)]^{3/2}$ as expected. But in some cases it could be different. The dependence of $\phi(s,P)$ has a minimum at $s = s_{min}$ and a maximum at $s = s_{max} \geq s_{min}$. As $T \to T_i$ $s_{min} \to s_{max}$ and $\Delta\phi \to 0$. At $T = T_i$ there is no barrier, minimum and maximum merge, forming an inflection point. Let us expand $\phi(s,P)$ in power series in $(s - s_i)$ and consider the first two non-vanishing terms of the series:

$$\phi(s,P) = \phi(s_i,P) + [(\partial h/\partial s)_{s_i} - T](s - s_i) - [T_i/(2n + 1)Q^{2n}](s - s_i)^{2n+1}, \tag{54}$$

where $(\partial h/\partial s)_{s_i} = T_i$ as was shown above, $n$ is a positive integer, and $Q$ is some coefficient. Then the condition $\partial\phi(s,P)/\partial s = 0$ yields

$$s_m = s_i \pm Q[1 - (T/T_i)]^{1/2n}. \tag{55}$$

Eqs (54) and (55) yield [9]

$$\Delta\phi(T) = [4nQT_i/(2n + 1)][1 - (T/T_i)]^{(2n+1)/2n}. \tag{56}$$

Eq. (56) shows that only when $n = 1$ [the right-hand part of Eq. (54) has a linear and a cubic terms only], the barrier vanishes as $[1 - (T/T_i)]^{3/2}$ at $T \to T_i$. In general case the temperature dependence of the barrier height, $\Delta\phi(T)$, could be different [9].

In summary it should be said that the application of the proposed approach yields results consistent with standard predictions from thermodynamic stability theory and allows one to consider within the framework of the same formalism a series of thermodynamic problems.

COMMENT

Theory of alloys is usually formulated in the framework of a *configurational model*. In this model a contribution of temperature is neglected, while calculating e*nergy* and *entropy*. This is justified because the contribution of temperature is many orders of magnitude smaller than the contribution of interatomic interactions and atomic distributions.

In the *Theory of Alloys energy* and *entropy* are usually calculated in the framework of a *configurational* model as functions of the *order parameter*, $\eta$. Then the *Gibbs free energy (GFE)*, $G(\eta)$, is extremised on the order parameter. Maxima and minima are usually found. Low maxima reduce stability. Some of the minima corresponds to stable states, the other ones to the metastable states (if the dividing GFE barrier, the difference between the corresponding maximum and minimum of the GFE, is comparatively low). Different phases, different order parameters, which could be defined in different ways. So we have: $G(\eta) = H(\eta) - S(\eta)T$, here $H(\eta)$ is the Enthalpy.

Once I've thought: why not to choose a parameter to minimize the GFE on as $S$ itself instead of $\eta$? As a rule $S$ is a monotonous function of $\eta$, so it is ok as a rule. *It's just a matter of transformation of variables, of substitution of one variable with another one*. Then it is possible to formulate a *general approach* for the description of metastable states, on the basis of the following equation: $G(S,P) = H(S,P) - ST$ [by the way, it is well known in *Thermodynamics* that $H$ is defined as $H(S,P)$]. These considerations had helped me to formulate my more general approach.